\documentstyle[11pt,amssymb,epsf]{article}

\def\ben{\begin{equation}}
\def\een{\end{equation}}

  \let\n=\nu  \let\p=\pi

\let\C=\Chi

\def\nn{\nonumber} \def\bd{\begin{document}} \def\ed{\end{document}}
\def\ds{\documentstyle} \let\fr=\frac \let\bl=\bigl \let\br=\bigr
\let\Br=\Bigr \let\Bl=\Bigl
\let\bm=\bibitem
\let\na=\nabla
\let\pa=\partial \let\ov=\overline
\newcommand{\be}{\begin{equation}}
\newcommand{\ee}{\end{equation}}
\def\ba{\begin{array}}
\def\ea{\end{array}}
\def\ft#1#2{{\textstyle{{\scriptstyle #1}\over {\scriptstyle #2}}}}
\def\fft#1#2{{#1 \over #2}}
\def\del{\partial}
\def\vp{\varphi}
\def\sst#1{{\scriptscriptstyle #1}}
\def\oneone{\rlap 1\mkern4mu{\rm l}}
\def\td{\tilde}
\def\wtd{\widetilde}
\def\ie{\rm i.e.\ }
\def\dalemb#1#2{{\vbox{\hrule height .#2pt
        \hbox{\vrule width.#2pt height#1pt \kern#1pt
                \vrule width.#2pt}
        \hrule height.#2pt}}}
\def\square{\mathord{\dalemb{6.8}{7}\hbox{\hskip1pt}}}
\newcommand{\ho}[1]{$\, ^{#1}$}
\newcommand{\hoch}[1]{$\, ^{#1}$}
\newcommand{\bea}{\begin{eqnarray}}
\newcommand{\eea}{\end{eqnarray}}
\newcommand{\ra}{\rightarrow}
\newcommand{\lra}{\longrightarrow}
\newcommand{\Lra}{\Leftrightarrow}
\newcommand{\bp}{\tilde \beta^\prime}
\newcommand{\tr}{{\rm tr} }
\newcommand{\Tr}{{\rm Tr} }
\def\0{{\sst{(0)}}}
\def\1{{\sst{(1)}}}
\def\2{{\sst{(2)}}}
\def\3{{\sst{(3)}}}
\def\4{{\sst{(4)}}}
\def\5{{\sst{(5)}}}
\def\6{{\sst{(6)}}}
\def\7{{\sst{(7)}}}
\def\8{{\sst{(8)}}}
\def\n{{\sst{(n)}}}
\def\cA{{{\cal A}}}
\def\cB{{{\cal B}}}
\def\cF{{{\cal F}}}
\def\cH{{{\cal H}}}
\def\tV{\widetilde V}
\def\tW{\widetilde W}
\def\tH{\widetilde H}
\def\tE{\widetilde E}
\def\tF{\widetilde F}
\def\tA{\widetilde A}
\def\im{{i}}
\def\tY{{{\wtd Y}}}
\def\ep{{\epsilon}}
\def\vep{{\varepsilon}}
\def\R{\rlap{\rm I}\mkern3mu{\rm R}}
\def\bD{{{\bar D}}}

\def\R{\rlap{\rm I}\mkern3mu{\rm R}}
\def\bD{{{\bar D}}}
\def\R{{{\Bbb R}}}
\def\C{{{\Bbb C}}}
\def\H{{{\Bbb H}}}
\def\CP{{{\Bbb C}{\Bbb P}}}
\def\RP{{{\Bbb R}{\Bbb P}}}
\def\Z{{{\Bbb Z}}}
\def\bA{{{\Bbb A}}}
\def\bB{{{\Bbb B}}}
\def\bC{{{\Bbb C}}}
\def\bD{{{\Bbb D}}}
\def\bE{{{\Bbb E}}}
\def\bZ{{{\Bbb Z}}}
\def\Re{{{\frak{Re}}}}
\def\Im{{{\frak{Im}}}}
\def\cosec{{\,\hbox{cosec}\,}}
\def\Gm{{\Gamma_{\!\! -}}}
\def\Gp{{\Gamma_{\!\! +}}}
\def\stan{{standard }}
\def\nonstan{{supernumerary }}
\def\p{{\partial}}

\def\bog{{Bogomol'nyi\ }}

\newcommand{\tamphys}{\it Center for Theoretical Physics,
Texas A\&M University, College Station, TX 77843}

\newcommand{\upenn}{\it Department of Physics and Astronomy,\\ University
of Pennsylvania, Philadelphia, PA 19104}

\newcommand{\brussels}{\it Physique Th\'eorique et Math\'ematique,
Universit\'e Libre de Bruxelles,\\ Campus Plaine C.P. 231, B-1050
Bruxelles, Belgium}

\newcommand{\damtp}{\it DAMTP, Centre for Mathematical Sciences,\\
 Cambridge University, Wilberforce Road, Cambridge CB3 OWA, UK}

\newcommand{\auth}{Z.W. Chong\hoch{\ddagger}, M. Cveti\v c\hoch{*2},   
H. L\"u\hoch{\ddagger1} and C.N. Pope\hoch{\ddagger1}}

\begin{document}

\begin{flushright}
MIFP-06-17\ \ \ UPR-1156-T \\
{\bf hep-th/0606213}\\
June\  2006
\end{flushright}

\vspace{10pt}

\begin{center}

{\large {\bf Non-Extremal Rotating Black Holes in Five-Dimensional 
Gauged Supergravity}}

\vspace{20pt}
\auth

\vspace{10pt}{\hoch{*}\it Department of Physics and Astronomy,\\
University of Pennsylvania, Philadelphia, PA 19104, USA}

\vspace{10pt}{\hoch{\ddagger}\it George P. \& Cynthia W. Mitchell
Institute for Fundamental Physics,\\ Texas A\& M University,
College Station, TX 77843-4242, USA}


%
%
%

\vspace{20pt}


\begin{abstract}

   Supersymmetric black holes in five-dimensional gauged supergravity must
necessarily be rotating, and so in order to study the passage to black holes
away from supersymmetry, it is of great interest to obtain non-extremal
black holes that again have non-zero rotation. In this paper we find a
simple framework for describing
non-extremal rotating black holes in five-dimensional gauged
supergravities.  Using this framework, we are able to construct a new
solution, describing the general single-charge solution of ${\cal N}=2$
gauged supergravity, with arbitrary values for the two rotation parameters. 
Previously-obtained solutions with two or three equal charges also assume
a much simpler form in the new framework, as also does the general 
solution with three unequal charges in ungauged ${\cal N}=2$ supergravity.
We discuss the thermodynamics and BPS limit of the new single-charge 
solutions, and we discuss the separability of the Hamilton-Jacobi and 
Klein-Gordan equations in these backgrounds.

\end{abstract}
\end{center}

{\vfill\leftline{}\vfill \vskip 10pt \footnoterule {\footnotesize
{\footnotesize
\hoch{1} Research supported in part by DOE grant
DE-FG03-95ER40917.}\vskip 2pt
\hoch{2} Research supported in part by DOE grant
DE-FG02-95ER40893, NSF grant INTO3-24081, and the\\
$\phantom{xxxxi}$  Fay R. and Eugene L.
Langberg Chair.}\vskip 2pt
}

\pagebreak

\newpage

\section{Introduction}

   With the developments in the understanding the AdS/CFT correspondence 
over the last few years, the investigation of solutions of gauged
supergravities, and, especially, five-dimensional gauged supergravity,
has acquired a new significance.  In particular, it is of considerable
interest to study five-dimensional solutions that provide a generalisation
of the pure anti-de Sitter geometry whose dual boundary theory is an
${\cal N}=4$ supersymmetric conformal field theory.  Black holes provide
a natural class of such generalisations.  As was discussed in 
\cite{hawhuntay}, by considering rotating asymptotically AdS black holes, 
one can make contact via the AdS/CFT correspondence with four-dimensional 
conformal field theories in a rotating Einstein universe.

   The example considered in \cite{hawhuntay} was the purely gravitational
Kerr-AdS black hole in five dimensions.   This is a non-supersymmetric 
background, and so from the AdS/CFT viewpoint it does not enjoy the 
the added control and protection that would be exhibited in a BPS or 
near-BPS configuration.   Supersymmetric rotating black holes have been 
found in five-dimensional gauged supergravities 
\cite{gutrea1,gutrea2,chcvlupo1,chcvlupo2,kulure}.  It is worth emphasising
that there are no static black-hole solutions in five-dimensional
gauged supergravity, and that rotation is essential in order to avoid
the occurrence of naked singularities.  Perhaps the 
greatest interest
from the point of view of the dual CFT's arises if one can study
AdS black hole backgrounds as one moves away from the supersymmetric situation,
and this provides a further motivation for studying non-extremal black
holes with rotation.

   In the case of five-dimensional minimal gauged supergravity, the
general solution, with mass parameter $M$, charge parameter $q$ and
the two independent rotation parameters $a$ and $b$, was obtained
in \cite{chcvlupo2}.  This generalised a previous result in \cite{cvlupo1},
in which the two rotation parameters were set equal. 

   It would also be of interest to obtain the general non-extremal rotating
solution in the case of five-dimensional ${\cal N}=2$ supergravity coupled
to matter.  The most relevant case corresponds to ${\cal N}=2$ supergravity
coupled to two vector multiplets.  This case, which therefore has
three vector fields in total, gauging $U(1)^3$, can also be viewed as the
truncation to the abelian subgroup of the maximal gauged $SO(6)$ 
supergravity.  The solution for non-extremal rotating black holes 
with $a=b$ was obtained in \cite{cvlupo2}.  A special case in which two
of the three charges are set equal, with the third having a value
related to these, was obtained for all $a\ne b$ in \cite{chcvlupo1}.  Also
in the same paper, another special case, with $b=0$ and only one charge,
was constructed.

   In the present paper, we extend this previous single-charge 
result by obtaining the
general solution for a five-dimensional rotating black hole with 
arbitrary rotation parameters $a$ and $b$, in the case that just one of
the three $U(1)$ gauge fields carries a charge.  In some sense this can 
be viewed as the most general ``basic'' solution of the ${\cal N}=2$
theory.  Our approach to 
constructing this solution involves first recasting the metrics into a 
form that leads eventually to a rather simple presentation of the 
result.  We also find that the same type of transformation, applied to
previously-known cases, leads to rather simple expressions in those
cases too.

   The paper is organised as follows.  In section 2 we introduce our
general ansatz for the class of rotating black-hole metrics we shall 
consider.  Then, we present our new results for the general single-charge 
rotating 
black holes.  In section 3 we discuss the thermodynamics of the new
solutions, followed in section 4 by the construction of the supersymmetric 
limit.  In section 5, we examine the previously-known rotating 
black hole solutions in five-dimensional gauged
supergravity, and show how these too fit elegantly within the 
formulation that we have adopted in this paper.
In addition, we show that the general 3-charge solution in ungauged
five-dimensional supergravity, which was constructed in \cite{cvetyoum},
also has a very simple expression when written in this formalism.
The paper ends with conclusions in section 6.

\section{The General Single-Charge Rotating Black Hole}\label{formsec}

   In this section, we construct the general solution for a single-charge
non-extremal rotating black hole in five-dimensional $U(1)^3$ gauged
${\cal N}=2$ supergravity, characterised by the mass, charge, 
and two rotation parameters $a$ and $b$.  

    The bosonic sector of the relevant ${\cal N}=2$ theory can be 
derived from the Lagrangian 
\be
e^{-1}\, {\cal L} = R - \ft12{\del\vec\varphi}^2 -
  \ft14\sum_{i=1}^3 X_i^{-2}\, {(F^i)}^2  + 4 g^2 \,
  \sum_{i=1}^3 X_i^{-1} + \ft1{24} |\ep_{ijk}|\, \ep^{\mu\nu\rho\sigma\lambda}
  F^i_{\mu\nu}\, F^j_{\rho\sigma}\, A^k_{\lambda}\,,\label{d5lag}
\ee
where $\vec\varphi=(\varphi_1,\varphi_2)$, and
\be
X_1= e^{-\fft1{\sqrt6}\varphi_1 -\fft1{\sqrt2} \varphi_2}\,,\qquad
X_2= e^{-\fft1{\sqrt6}\varphi_1 +\fft1{\sqrt2} \varphi_2}\,,\qquad
X_3 = e^{\fft2{\sqrt6}\varphi_1}\,.
\ee
All the solutions that we shall consider in this paper, comprising the
new general single-charge rotating black holes, and also the
previously-known solutions with two equal charges or three equal charges,
as well as the general solutions in the ungauged theory with three unequal
charges, can all be cast in a simple manner within the following
formalism.  We write the metrics as
\bea
ds_5^2 &=& (H_1 H_2 H_3)^{1/3}\, (x+y)\, d\hat s_5^2\,,\nn\\
d\hat s_5^2 &=& -\Phi\, (dt + {\cal A})^2 + ds_4^2\,,\label{kkmet}
\eea
with the scalars and gauge potentials given by
\bea
X_i &=& H_i^{-1}\, (H_1 H_2 H_3)^{1/3}\,,\nn\\
A^1 &=& \fft{2m}{x+y}\, H_1^{-1}\, \{s_1 c_1 dt + s_1 c_2 c_3 [ab d\chi 
     + (y-a^2-b^2) d\sigma] + c_1 s_2 s_3 (ab d\sigma - y d\chi)\}\,,
\,,\label{XAi}
\eea
with $A^2$ and $A^3$ given by cyclically permuting the subscripts on the
right-hand side.  The functions $H_i$ are given by
\be
H_i = 1 + \fft{2m s_i^2}{x+y}\,,
\ee
and we are using the shorthand notation 
\be
s_i= \sinh\delta_i\,,\qquad c_i = \cosh\delta_i\,,
\ee
where $\delta_i$ are the charge parameters.  The four-dimensional base
metric in (\ref{kkmet}) takes the form
\be
ds_4^2 = \Big( \fft{dx^2}{4 X} + \fft{dy^2}{4Y}\Big) 
    + \fft{U}{G}\, \Big(d\chi - \fft{Z}{U}\, d\sigma\Big)^2 +
     \fft{X Y}{U}\, d\sigma^2\,,\label{4met}
\ee
where $X$ is a function of $x$, $Y$ is a function of $y$, and $G$, $U$
and $Z$ are functions of both $x$ and $y$.  The ``Kaluza-Klein'' 1-form
${\cal A}$ appearing in (\ref{kkmet}) lives purely in the four-dimensional
base space, and takes the form
\be
{\cal A} = f_1\, d\sigma + f_2\, d\chi\,.
\ee
The functions $f_1$ and $f_2$ depend only on $x$ and $y$, as does 
$\Phi$, which is given by
\be
\Phi = \fft{G}{(x+y)^3\, H_1\, H_2\, H_3}\,.
\ee
The inverse of the metric $d\hat s_5^2$ is given by 
\bea
\Big(\fft{\del}{\del\hat s_5}\Big)^2 &=& -\fft1{\Phi}\,
\Big(\fft{\del}{\del t}\Big)^2 + 4 X\, \Big(\fft{\del}{\del x}\Big)^2 +
               4 Y\, \Big(\fft{\del}{\del y}\Big)^2+ 
  \fft{G}{U}\, \Big(\fft{\del}{\del\chi} 
    - f_2 \, \fft{\del}{\del t}\Big)^2 \nn\\
&&+
  \fft1{UXY}\, \Big( U\fft{\del}{\del\sigma} + 
            Z\, \fft{\del}{\del\chi} - (f_1\, U + f_2\, Z)\, 
                      \fft{\del}{\del t}\Big)^2 \,.\label{invhat}
\eea

   Since there is no solution-generating technique for deriving charged
black holes from neutral black holes in gauged supergravity (unlike
the situation in ungauged supergravity), there is really no way other
than a combination of guesswork, followed by explicit verification,
for obtaining the charged solutions.  We were led to write the ansatz
for the metric, gauge potentials and scalar fields in the manner we have
presented above by considering all the previously-obtained examples.  The
specific results for the new general single-charged rotating black holes,
which we shall present below, were obtained by making a detailed
comparison of various known cases, transformed into the format of the
ansatz above, and then making a conjecture for the form of the solution.
Finally, we substituted this into the equations of motion following
from (\ref{d5lag}), to verify that it was indeed a solution.  In doing
this, we made extensive use of the Mathematica algebraic computing language.

    Our new results for the general single-charge rotating black hole in
five-dimensional gauged ${\cal N}=2$ supergravity are as follows.  Taking
$\delta_2=\delta_3=0$, and writing $\delta_1=\delta$, we find
\bea
X &=& (x+a^2)(x+b^2)- 2mx + g^2(x+a^2)(x+b^2)[x+2ms^2 -(a^2+b^2) s^2 + 2absc]
\,,\nn\\
Y&=& - (a^2-y)(b^2-y)[1 - g^2(y+(a^2+b^2)s^2 -2absc)]\,,\nn\\
G &=& (x+y)(x+y-2m) + g^2(x+y)^2\, (x-y+a^2+b^2)H \,,\nn\\
U &=& y X - x Y + s^2 W\,,\qquad Z = a b (X+Y) + s c W\,,\nn\\
W &=& - 2g^2 m (a^2-y)(b^2-y) x + g^4(x+a^2)(x+b^2)(a^2-y)(b^2-y)(x+y+2ms^2)
\,,\nn\\
\Phi &=& \fft{G}{(x+y)^3\, H}\,,\nn\\
{\cal A}&=& s\, (x d\chi + abd\sigma) + 
           c\,[ab d\chi -(x+a^2+b^2)d\sigma]\nn\\
&& + \fft1{G}\, \Big[-s\, (x+y-2m)(xd\chi + abd\sigma) - c\, (x+y)
          [ab d\chi -(x+a^2+b^2-2m)d\sigma] \nn\\
&&\qquad\quad 
  + g^2(x+a^2)(x+b^2)(x+y+2ms^2)(c d\sigma- s d\chi)\Big]\,.
\eea
The gauge potentials in (\ref{XAi})  reduce to $A^2=A^3=0$
and
\be
A^1= \fft{2ms}{x+y+2ms^2}\, [c dt + ab d\chi + (y-a^2-b^2)d\sigma]\,,
\ee
and the $H_i$ functions are given by $H_2=H_3=1$ and
\be
H_1 \equiv  H= 1 + \fft{2ms^2}{x+y}\,.
\ee
 
   The solution we have presented here has four non-trivial
parameters, namely $m$, $\delta$, $a$ and $b$ (with $s=\sinh\delta$, 
$c=\cosh\delta$), which characterise the mass, charge and two angular
momenta respectively.  In the next section, we shall derive the 
conserved charges associated with this solution, and also study the
thermodynamic quantities.

\section{Global Structure and Thermodynamics}

     The black hole solution we have constructed is of cohomogeneity 2, 
with the metric functions depending on the
 non-compact radial coordinate $x$, and the compact coordinate
$y$ which runs from $y=a^2$ to $y=b^2$.  If we define
$y=a^2 \cos^2\theta + b^2\sin^2\theta$, then $\theta$ plays the role of
the latitude
coordinate of the round 3-sphere, viewed as a foliation of Clifford tori,
\be
d\Omega_3^2=d\theta^2 + \cos^2\theta\,d\phi^2 + \sin^2\theta\,
d\psi^2\,.
\ee

   The coordinates $\sigma$ and $\chi$ can be related to  
two azimuthal $U(1)$ coordinates $\phi$ and $\psi$ with canonical 
$2\pi$ periods via the redefinitions
\be
\sigma = \fft{1}{a^2-b^2} \Big( \fft{a(\phi-\td a\, g^2\,t)}
{\Xi_{\td a}} - \fft{b(\psi-\td b\, g^2\,t)}{\Xi_{\td b}} \Big)
\,,\qquad \chi = \fft{1}{a^2-b^2} \Big(\fft{b(\phi-\td a\, g^2\,t)}
{\Xi_{\td a}} -  \fft{a(\psi-\td b\, g^2\,t)}{\Xi_{\td b}} \Big)\,,
\ee
where
\be
\Xi_{\td a} =1 - \td a^2 g\,,\qquad
\Xi_{\td b} =1 - \td b^2 g^2\,,
\ee
and we have defined ``boosted'' rotation parameters $\td a$ and $\td b$, 
given by
\be
\td a = a\, c - b\, s\,,\qquad \td b = b\, c - a\, s\,.
\ee
The inclusion of $t$ in the redefinitions ensures that $\phi$ and $\psi$
define an asymptotically static frame.

The radial coordinate $x$ runs to the asymptotically flat region 
as $x\rightarrow\infty$, and there is an outer horizon at $x=x_0$, the
largest root of $X$.   Straightforward calculations show that
the entropy and temperature are given by
\bea
S &=& \fft{\pi^2 (x_0 + a^2)(x_0+b^2) (c + s\, \td a\, \td b\, g^2)}
{2\Xi_{\td a} \Xi_{\td b}\sqrt{ x_0 - g^2 s^2 (x_0+a^2)(x_0 + b^2)}}
\,,\nn\\
T &=& \fft{\pi}{4S\, \Xi_{\td a}\Xi_{\td b} [x_0 - g^2 s^2
(x_0+a^2)(x_0 + b^2)]}\Big(x_0^2 - a^2 b^2 -g^4 s^2
(x_0+a^2)^2 (x_0 + b^2)^2\nn\\
&&\qquad + g^2[x_0^2 (2x_0 + a^2 + b^2) - s(x_0^2-a^2b^2)(
2a\,b\,c - s(a^2 + b^2)]\Big)\,.
\eea

   The angular velocities of the horizon, measured with respect to
the azimuthal coordinates $\phi$ and
$\psi$ of the asymptotically static frame, are given by
\be
\Omega_{\phi}= \fft{a (c+s) (\Xi_{\td a} + g^2 (x_0 + a^2))}
{(x_0 + a^2)(1 + s (c+s) (1 + \td a\, \td b\, g^2)}\,,\qquad
\Omega_{\psi} = \fft{b (c+s) (\Xi_{\td b} + g^2 (x_0 + b^2))}
{(x_0 + b^2)(1 + s (c+s) (1 + \td a\, \td b\, g^2)}\,.
\ee
The corresponding angular momenta, defined via Komar 
integrals $J=1/(16\pi)
 \int{*dK}$ where $K=\del/\del\phi$ or $\del/\del\psi$, are given by
\be
J_\phi= \fft{\pi\, a\, m (c + s\, \td a\, \td b\, g^2)}
{2\Xi_{\td a}^2 \,\Xi_{\td b}}\,,\qquad
J_\phi= \fft{\pi\, b\, m (c + s\, \td a\, \td b\, g^2)}
{2\Xi_{\td a} \, \Xi_{\td b}^2}\,.
\ee
The electric potential and conserved electric charge are given by
\bea
\Phi&=&\fft{s(c+s)(\Xi_{\td a} + g^2 (x_0 + a^2))}
{1 + s (c+s)(1 + \td a\, \td b\, g^2)} =
\fft{s(c+s)(\Xi_{\td b} + g^2 (x_0 + b^2))}
{1 + s (c+s)(1 + \td a\, \td b\, g^2)}\,,\nn\\
Q&=&\fft{\pi m\, s(c + s\,\td a\, \td b\, g^2)}{2\Xi_{\td a}\, \Xi_{\td b}}
\,.
\eea
Note that we have the identity $\td \Xi_{\td a} + g^2 (x_0 + a^2)=
\td \Xi_{\td b} + g^2 (x_0 + b^2)$, which is implied by the relation
 $a^2-b^2=\td a^2 -\td b^2$.

    The mass (or energy) of the black hole can be easily obtained by 
integrating the first law of thermodynamics,  $dE=T\, dS + \Omega_\phi\,
dJ_\phi + \Omega_\psi\, dJ_\psi + \Phi\,dQ$; it is given by
\be
E=\fft{\pi\,m}{4\Xi_{\td a}^2\, \Xi_{\td b}^2}
\Big\{2 \Xi_{\td a} + 2 \Xi_{\td b} - \Xi_{\td a}\Xi_{\td b}
+ (\Xi_{\td a} + \Xi_{\td b}) [4s\,c\,\sqrt{(1-\Xi_{\td a})(
1-\Xi_{\td b})} + s^2 (2-\Xi_{\td a})(2-\Xi_{\td b})]\Big\}
\,,
\ee

\section{The BPS limit}

    By considering the AdS superalgebra, as was discussed in 
\cite{cvgilupotime}, one can see that a BPS limit of the general 
non-extremal solution arises when 
\be
E + g\, J_\phi + g\, J_\psi + Q =0\,.\label{bps}
\ee
(Equivalent BPS limits arise for other sign choices.)
From our expressions in the previous section, we find that (\ref{bps}) 
implies
\be
e^{2\delta} = 1 + \fft{2}{(\td a + \td b) g}\,.
\ee
Expressed in terms of the original rotation parameters $a$ and $b$, this
BPS condition can be rewritten as
\be
(a+b)g\, \sinh\delta=1\,.
\ee

   In this supersymmetric limit, where there exists a Killing
spinor $\eta$, the Killing vector
\be
\ell = \fft{\del}{\del t} - g \fft{\del}{\del \phi} -
g \fft{\del}{\del \psi}
\ee
has a spinorial square root, in the sense that $\ell^\mu= \im\, \bar\eta\, 
 \gamma^\mu\eta$.  It has norm given by
\be
\ell^\mu\, \ell_\mu = - \fft{1}{H}\,.
\ee
As discussed in \cite{gutrea1}, the single-charge BPS rotating solution with
a single angular momentum always has naked closed timelike curves, 
and the inclusion of the additional rotation parameter 
does not alter the feature.  Thus there are no regular black hole
solutions in the BPS limit.

\section{Previously-Known Rotating Black Holes}

   In this section, we present the previously-known rotating black
hole solutions of five-dimensional supergravity, using the formalism that
we have introduced in section \ref{formsec}.  These amount to three
cases.  The first is the case found in \cite{chcvlupo1} with two charges
set equal and the third related to this, in gauged ${\cal N}=2$ supergravity.
The second case, obtained in \cite{chcvlupo2}, is where all three charges 
are equal in ${\cal N}=2$ gauged supergravity; this can be viewed also 
as the general solution in minimal gauged supergravity.  The third case
is the general solution in ungauged ${\cal N}=2$ supergravity, 
with three unequal charges, which was obtained in \cite{cvetyoum}.  All 
three of these cases can be represented elegantly within the formulation
of section \ref{formsec}, and thus to present them we need only specify
the various functions and gauge potentials.

\subsection{Two equal charges in gauged supergravity}

   In this solution, obtained in \cite{chcvlupo1}, we have 
$\delta_1=\delta_2=\delta$, with $\delta_3=0$.  In the ungauged theory, this
choice of charge parameters would imply that two of the three physical 
conserved charges were equal and non-vanishing, whilst the third 
vanished.  As was
shown in \cite{chcvlupo1}, in the case of the solution in gauged supergravity
the third physical charge is actually non-vanishing too, with a value 
related to those of the other two.  We find that in the formalism of
section \ref{formsec}, this solution is given by  
\bea
X &=& (x+a^2)(x+b^2)- 2mx + g^2(x+a^2+2ms^2)(x+b^2+2ms^2) x \,,\nn\\
Y&=& - (a^2-y)(b^2-y)(1 - g^2\, y)\,,\nn\\
G &=& (x+y)(x+y-2m) + g^2(x+y)^2\, (x-y+a^2+b^2 + 2ms^2) H\,,\nn\\
U &=& y X - x Y \,,\qquad Z = a b (X+Y) \,,\nn\\
\Phi &=& \fft{G}{(x+y)^3\,  H^2}\,,\nn\\
{\cal A}&=& ab d\chi - (x+a^2+b^2 +2ms^2)d\sigma \nn\\
&& + \fft1{G}\, \Big[ -(x+y+2ms^2)[ab d\chi - (x+a^2+b^2 -2m)d\sigma]
    \nn\\
&&\qquad + g^2(x+a^2+2ms^2)(x+b^2+2ms^2)(x+y+2ms^2)d\sigma\Big] \,.
\eea
That the gauge potentials in (\ref{XAi}) reduce to 
\bea
A^1&=&A^2 =\fft{2msc}{x+y+2ms^2}\, [dt + ab d\chi + (y-a^2-b^2)d\sigma]\,,\nn\\
A^3 &=& \fft{2ms^2}{x+y}\, (ab d\sigma - y d\chi)\,,
\eea
and the functions $H_i$ reduce to $H_3=1$, and 
\be
H_1=H_2=H=1 + \fft{2ms^2}{x+y}\,.
\ee

\subsection{Three equal charges in gauged supergravity}

   This solution, obtained in \cite{chcvlupo2}, 
which can also be viewed as the general rotating 
black hole solution in five-dimensional minimal gauged supergravity,
corresponds in the formalism of section \ref{formsec} to taking
$\delta_1=\delta_2=\delta_3=\delta$.  We find that it then takes the 
form
\bea
X &=& (x+a^2)(x+b^2)- 2mx \nn\\
&&+ g^2(x+a^2+ 2ms^2)(x+b^2+2ms^2)
       [x+2ms^2 -(a^2+b^2) s^2 + 2absc]
\,,\nn\\
Y&=& - (a^2-y)(b^2-y)[1- g^2(y+(a^2+b^2)s^2 -2absc)]\,,\nn\\
G &=& (x+y)(x+y-2m) + g^2(x+y)^2\,  (x-y+a^2+b^2 + 2ms^2) H^2\,,\nn\\
U &=& y X - x Y + s^2 W\,,\qquad Z = a b (X+Y) + s c W\,,\nn\\
W &=& - 2g^2 m (a^2-y)(b^2-y) [x (c^2+s^2) + (a^2+b^2) s^2 + 2m s^4]\nn\\
&& + g^4(x+a^2+2ms^2)(x+b^2+2ms^2)(a^2-y)(b^2-y)(x+y+2ms^2)
\,,\nn\\
\Phi &=& \fft{G}{(x+y)^3\,  H^3}\,, \\
{\cal A}&=& s (x d\chi + ab d\sigma) + c[ab d\chi - (x+a^2+b^2+ 2ms^2) d\sigma]
 + 2m s^3\, d\chi  \nn\\
&& + \fft{H}{G}\, \Big[ - s (x+y-2m)(xd\chi + ab d\sigma) -
  c (x+y)[ab d\chi - (x+a^2+b^2- 2m) d\sigma]\nn\\
  && + g^2(x+a^2+2ms^2)(x+b^2+2ms^2)(x+y+2ms^2)
               (cd\sigma -s d\chi)\Big]\,.\nn 
\eea
The gauge potentials in (\ref{XAi}) reduce to 
\be
A^1=A^2=A^3= \fft{2msc}{x+y+2ms^2}\, 
  \{dt + c[ab d\chi + (y-a^2-b^2)d\sigma] + s(abd\sigma-yd\chi) \}\,,
\ee
and the functions $H_i$ are given by
\be
H_1=H_2=H_3=H\equiv 1 + \fft{2ms^2}{x+y}\,.
\ee

\subsection{Three unequal charges in ungauged supergravity}

   This solution was first obtained in \cite{cvetyoum}, by applying
a solution-generating procedure to add charges to the neutral
five-dimensional rotating black hole of Myers and Perry \cite{myeper}. 
We find that in the formulation of section \ref{formsec}, it takes the
simple form
\bea
X &=& (x+a^2)(x+b^2)- 2mx \,,\nn\\
Y&=& - (a^2-y)(b^2-y) \,,\nn\\
G &=& (x+y)(x+y-2m)  \,,\nn\\
U &=& y X - x Y \,,\qquad Z = a b (X+Y) \,,\nn\\
\Phi &=& \fft{G}{(x+y)^3\,  H_1 H_2 H_3}\,,\nn\\
{\cal A}&=& \fft{2m c_1 c_2 c_3}{G} [(a^2+b^2-y) d\sigma -a b d\chi]
-
      \fft{2m s_1 s_2 s_3}{x+y}\, (ab d\sigma - y d\chi)\,.
\eea

\section{Conclusions}

    In this paper, we have constructed the general solution for a 
singly-charged rotating black hole in five-dimensional gauged supergravity.  
It can be viewed as a solution in the maximal ${\cal N}=8$ gauged theory,
with the charge being carried in a single $U(1)$ subgroup of the $SO(6)$
gauge group.  Equivalently, it can be viewed as a solution in ${\cal N}=2$
gauged supergravity coupled to two vector multiplets, with the charge carried
in a single $U(1)$ factor of the $U(1)^3$ gauge group.  The solution
generalises a special case obtained in \cite{chcvlupo1}, in which
only one of the two rotation parameters was non-vanishing.

   In addition to obtaining the new solution, we have also found a simple 
framework within which all the currently-known five-dimensional 
non-extremal rotating black holes can be described.  This framework 
involves writing the metric as a fibration over a four-dimensional base,
as given in (\ref{kkmet}) and (\ref{4met}).  The expressions for 
the new singly-charged solution appeared in section 2, and for the 
previously-obtained solutions with two or three equal charges in section 5.
Also in section 5, we presented the strikingly-simple expressions in this 
framework for 
the ungauged rotating  black holes with three unequal charges.  One may 
hope that the rather simple forms of all these examples may persist in
the most general case with three unequal charges in gauged supergravity,
which is not yet known explicitly.

   It is interesting to note that all the known five-dimensional 
rotating black hole metrics discussed in this paper have the feature that
the Hamilton-Jacobi equation and the Klein-Gordon equation in these 
backgrounds exhibit separability.  To be precise, the massless 
Hamilton-Jacobi and Klein-Gordon equations are separable in all
the backgrounds, and additionally, the massive equations are also
separable if the three charges are equal. 
The determinant of the metric is given by 
$\sqrt{-g_5} = \ft14 (H_1 H_2 H_3)^{1/3}\, (x+y)$, and hence 
\be
\sqrt{-g_5}\, \Big(\fft{\del}{\del s_5}\Big)^2 = \ft14 
        \Big(\fft{\del}{\del\hat s_5}\Big)^2 \,.
\ee
It is straightforward to check that the components of the hatted 
inverse metric, which can be read of
from (\ref{invhat}),  are all of the form of a sum of a function of $x$ and
a function of $y$ provided the charges are equal.  This immediately 
implies a manifest separability.  In the case of unequal charges one
has separability only if the mass is zero, implying that an overall
function of $x$ and $y$ can be factored out.  When the charges are equal,
the Hamilton-Jacobi and Klein-Gordon equations are separable both in the
massless and massive cases.

\section*{Acknowledgement}

   We are grateful to Paul Davis for drawing our attention to a
mis-statement about separability in an earlier version of this paper.
  M.C. is grateful to the Mitchell Institute for Fundamental Physics,
 H.L. and C.N.P. are grateful to the Physics Department at the
University of Pennsylvania, and 
C.N.P. is grateful to the DAMTP Relativity and Gravitation group, and
the Centre for Theoretical Cosmology at the CMS,
Cambridge, for hospitality during the course of this work.

\end{document}